# 'Interaction' versus 'action'
# in Luhmann's sociology of communication

In: Colin B. Grant (Ed.),
*Rethinking Interactive Communication:
New Interdisciplinary Horizons*,
Amsterdam: John Benjamins, forthcoming

Loet Leydesdorff
Science & Technology Dynamics, University of Amsterdam
Amsterdam School of Communications Research (ASCoR)
Kloveniersburgwal 48, 1012 CX  Amsterdam, The Netherlands
http://www.leydesdorff.net/index.htm ; loet@leydesdorff.net

## Abstract

Both 'actions' and 'interactions' can be considered as micro-operations that can be aggregated from a systemic perspective. Whereas actions operate historically, interactions provide the events retrospectively with meaning. Luhmann's sociology of communication systems adds to the approach of symbolic interactionism the question of what global dimensions of communication mean for local interactions. When communication is functionally differentiated—for example, in terms of media—tensions can be expected to develop between local organizations and global developments of communication structures. Interfaces enable us to translate selectively among (provisionally) stabilized representations, for example, in professional practices. 'Big science' and 'high tech' can be considered as organizational acculturations of an emerging level of sophistication in global communications. The global dimension remains a hypothesis, but entertaining this hypothesis of 'globalization' restructures the local expectations.

## Introduction

The theoretical oppositions between Luhmann (1984) and Habermas (1987) were framed in terms of 'systems theory' versus 'action theory' (cf. Habermas & Luhmann, 1971). 'Action theory' may seem less alienating than 'systems theory' because 'actions' can be intentional. The analytical distinction between theories that are based on 'action' or 'interaction' as micro-operations of social systems, however, is more fundamental than the one between action and systems theory. Luhmann's sociology can be considered as different from other systems-theoretical approaches because it assumes 'interaction' as the basic operation of social systems.

Both 'actions' and 'interactions' can be considered as micro-operations that can be aggregated from a systemic perspective. Actions can be aggregated, for example, into 'institutional agency,' whereas interactions may become increasingly complex by operating upon one another in a non-linear mode. Action can also be considered as an operation integrating social systems historically (Parsons, 1937; Habermas, 1981), while interactions may enable the actors to reproduce differentiation.

For example, the constructivist Latour (1987) proposed to 'follow the actors' in terms of their actions. Actions are then used as a historical *explanans*. The observation of an interaction, however, assumes a perspective from which one can reconstruct the observable events (e.g., actions). Like action, interaction occurs in history, but the system of reference for interaction is necessarily an interhuman construct. Interaction is by definition reflexive. The two operations of 'action' and 'interaction' cannot be reduced to each other because of this difference in their epistemological status.

From an interactive or network perspective, one can attribute an action to an actor, but this attribution can also be reconsidered. Interaction potentially rewrites the past, for example, from the perspective of a (historical) present. 'Interaction' thus provides us with an evolutionary category that operates at the network level, whereas actions remain to be attributed to the historical development of agency in terms of individuals or groups who carry the evolution of systems of social interactions. While actions can be expected to vary, interactions tend to evolve into systems of mutual expectations.

### The double hermeneutics of sociology

The epistemological difference between 'action' and 'interaction' was already fundamental to Weber's Marx-critique. Marx focused on historical action and wished to make predictions on this basis. Weber raised the question of the 'sociological meaning' of actions. From Weber's perspective, sociology uses historical instances for understanding the operation of analytical constructs (e.g., 'idealtypes'). Against Marx, Weber (e.g., 1904 and 1917) maintained that the historical accounts cannot inform us about a system's logic operating in history. The analyst 'understands' the actions in what can also be called a 'verstehende Soziologie.'

Understanding raises the question of how people and analysts construct meaning in *interactions*. A 'double hermeneutics' between the analyst's and the participant's level of action and accounting has since that time been a constitutive problem of sociology (Giddens, 1979). The dimension of external observation versus participation can be cross-tabled with the distinction between 'interaction' and 'action' (Table 1). A participant can also be an observer, but the analytical status of an observation is different from that of participation.

|  | *Participation* | *Observation* |
|---|---|---|
| *Action* | actor | report |
| *Interaction* | role | discourse |

**Table 1**
*The generation of a double hermeutics in sociology*

From the perspective of reflexive interaction at the network level, 'action' by a participant can be considered as a role attributed to or carried by an actor. Expectations with respect to actors are constructed within the network of communications among the observers. The interactive networks operate in terms of non-linear feedback loops on actions. From an action theoretical perspective, however, the network effects are attributed to the intentioned actors in terms of linear cause-effect relations. The unintended consequences of actions remain then unexplained.

By considering communication as the unit of analysis—or more precisely as 'the unit of operation'—of social systems, Luhmann's sociology shares with symbolic interactionism a focus on interaction. Symbolic interactionism with its roots in American pragmatism (Blumer, 1969), however, has been strongly contrasted to social systems theory (Grathoff, 1978). Luhmann mainly added to symbolic interactionism the question of what global dimensions of communication may mean for local interactions. How and to what extent are the local or 'first-order' observations structured by higher-order communications? But in order to ground the next-order level, Luhmann defined the basic operation of social systems as 'second-order observations': how does the network system enable us to make distinctions and to provide these distinctions with meaning at the network level?

## Symbolic Interactionism

In his authorative study of *symbolic interactionism,* Herbert Blumer (1969, at p. 8) stated:

> The importance lies in the fact that social interaction is a process that *forms* human conduct instead of being merely a means or a setting for the expression and release of human conduct.

Blumer traced the roots of the interactionist approach to George Herbert Mead's reformulation of the self as the result of a process of social interaction (Mead, 1934, at pp. 26f.). The communicative structure pervades action. Society, as Cooley (1902) once argued, exists inside the individual in the form of language and thought.

The basic unit of analysis in the interactionist account was defined as the joint act—the interactional episode (Lindesmith, Strauss, & Denzin, 1949, $^4$1975, at p. 4). The interactional episode is part of the larger society. In empirical studies, however, the larger social system was consistently treated as a result of interactions in micro-situations. Blumer (1969, at p. 58), for example, stated:

> However, in seeing the organization as an organization of actions symbolic interactionism takes a different approach. Instead of accounting for the activity of the organization and its parts in terms of organizational principles or system principles, it seeks explanation in the way the participants define, interpret, and meet the situations at their respective points. The linking together of this knowledge of concatenated actions yields a picture of the organized complex.

The resulting 'picture,' however, has the status of an account that can be communicated. This communication is no longer necessarily confined to the situation in which it emerged. As noted, the epistemological status of an account is different from an observable action because the observational report is reflexively organized. It contains a knowledge claim that can be validated by the participants and/or as a contribution to a sociology.

The need for a bottom-up approach to the validation does not follow logically from a focus on 'interactions,' but it was implied in the programmatic preference of symbolic interactionism for the analysis of 'micro-situations.' Knorr-Cetina (1981, at p. 27), for example, argued that the 'situational approach' is the road sociology must take for *methodological* reasons, since:

> (...) unlike the natural sciences the social sciences cannot hope to get to know the macro-order conceived in terms of emergent properties: they are methodologically bound to draw upon members' knowledge and accounts.

How can an analyst understand 'members' knowledge and accounts' other than by situating them in a context that has first to be (re)constructed from these same 'members' knowledge and accounts'? A reflexive turn is implied that adds to the analysts' understanding of the members' accounts. Whereas the micro-constructivists demand—as a methodological constraint—that the interpretation be validated locally, the accounts feed back into the situation from a perspective. This analytical angle makes the observation reflexively available as an observational report in contexts other than the ones in which they originated and were validated.

On the one hand, the micro-constructivists have substantiated their critique of systems approaches, arguing that in order to be useful for empirical research, a model should be able to account for the specificities of localized action and interaction. The focus on specific episodes has resulted in a richness of substantive understanding which cannot easily be brought into a systems perspective. The latter abstracts from the substance of the accounts by comparing them at the aggregated level. The reports can then be considered as contributions to a discourse. However, the accounts and not the actions reported within them are providing the variation from this perspective.

On the other hand, the situationalist approach fails us if we wish to understand why interactions are 'concatenated.' Some authors in this tradition have tried to specify control as, for example, 'alignment' (e.g., Fujimura, 1987), but the control mechanisms of the social system (e.g., codification processes) cannot fully be specified from within the situations. The historical report of the sequence only reflects the dynamics that produced the sequencing.

The systems perspective originates from taking a reflexive turn. Observations, for example, were defined by Luhmann (1984) from his second-order perspective as the operation of first distinguishing and then indicating the distinction made ("Unterscheiden und Bezeichnen"). The designation provides the distinctions with meaning. It should be noted that the operation of 'observation' thus defined implies two operations. By (re)combining the network operation with the historical information, the analytical perspective adds to understanding the historical cases. For example, one may also wish to raise the question why some things did *not* happen? In addition to the cases that happened to occur historically, one is sometimes—that is, under methodologically specifiable conditions—able to specify expectations about what might have happened. Historical accounts provide the systems theoretician with empirical materials for the formulation of hypotheses.

## Structuration theory

Some interactions are more likely to occur than others; previous interactions 'constrain and enable' future interactions. 'Structures' seem to operate as constraints both statically, that is, at each moment in time, and dynamically, that is, over time. Giddens (1979) proposed in his 'structuration theory' that structure be considered as a virtual operation which 'constrains and enables' action *ex ante* and 'aggregates' actions *ex post*. However, Giddens deliberately refrained from a specification of this 'duality of structure' as a virtual operation since—against Marxism and systems theory—the empirical sociologist should, in his opinion, foreground that 'social reproduction has itself to be explained in terms of the structurally bounded and contingently applied knowledgeability of social actors' (Giddens, 1981, at pp. 172 ff.).

In order to prevent any specification of structure 'outside time and space' in empirical research, Giddens (1984) then recommended *as a methodology* that structure be described only historically and contextually, that is, substantively operationalized in terms of historical instances. In Giddens's opinion, the mutual contingencies of structure and action can be studied by 'bracketing' the institutional dimension when the analysis is at the level of strategic conduct; and vice versa, the latter can be bracketed when one analyzes the former.[1] So, the two perspectives are developed as different views of the same matter; the two pictures together would provide a fuller insight into the mutual contingencies.

The definition of structure was thus shifted to 'a rule of sociological method,' but Giddens refused to draw the consequence of defining structure formally, that is, as a network operation. The 'virtual operation' of structure, however, is analytically different from its substantive instantiations. Giddens was himself aware of the problem that the core concept of his theory, that is, the 'duality of structure,' cannot be defined by using 'bracketing' (e.g., Giddens, 1979, at p. 95).

In my opinion, structuration theory contains all the elements, but for programmatic reasons, it denies the analyst the possibility of specifying the operation of 'structure' at the level of a social system. Furthermore, Giddens warned against making structure the subject of sociological theorizing when he strongly formulated (1984, at p. xxxvii):

> There can be no doubt about the sophistication and importance of the work of some authors currently endeavouring to develop Parsons's work in novel ways, particularly Luhmann and Habermas. But I think it is as necessary to repudiate the newer versions of Parsonianism as I do the longer established varieties of non-Parsonian structural sociology.

## Luhmann's proposal

Perhaps even more than Habermas, Luhmann has been deeply influenced by Parsons's systems theory. Parsons considered 'action' as the integrating operator of social systems: the analytical dimensions of an action are instantiated and reorganized in the performative dimension. The identification of the system with observable action, however, has led to a reification of systems-theoretical approaches. Social systems could then be considered as historical phenomena (Münch, 1982).

---

[1] Giddens (1979, p. 81) compares the concept of bracketing with *epoche* in the phenomenological tradition.

Although Parsons (1952) argued strongly that society should be considered as a category *sui generis*, from his perspective the social remains another dimension of an otherwise naturally given system. The social system was not further analyzed as an *interactive* and, therefore, cultural construct among human beings. Luhmann (1984) confronted this problem of confusing the historical level with the analytical by proposing to consider 'communication' as the running operator of the social system. Interaction can then be considered as a basic operation for producing meaning within social systems.

This proposal thoroughly solves the puzzle of combining the explanatory power of systems theory as a theory about communication and control with the richness of the descriptions in studies from interactionist traditions. The social system contains instances that were historically realized, but it can be considered as a multi-dimensional space of other options that could perhaps be realized in the future. The focus on the dynamics of the network enables us to integrate the micro- with the macro-approach. Middle-range approaches can also be appreciated because the analytical definition of the systems of reference becomes crucial to the specification of a research design.

Which networks can be considered as relevant for studying a specific research question? How can networks be delineated? Because of the freedom to specify expectations on analytical grounds, Luhmann's sociology is very different from those of Giddens or Habermas. The latter begin with historical observations, while Luhmann's theorizing begins with expectations that are based on 'horizons of meaning' (Luhmann, 2002b; cf. Husserl, 1954). This theory therefore allows for formalization without losing the relation to the interactive accounts. The interactive accounts provide the variation. From a network perspective, 'second-order observations' refer to a theory about possible observations. The observables can then be evaluated in relation to the theoretical expectations.

The social system is constructed bottom-up, but in a network mode. The interactions at the network level add uncertainty to the aggregations of lower-order units. In the formal language of statistics one can formulate that the aggregations contain 'within group' variation, but that one expects also 'between group' variation. A classical example is that of a school expected to contain more variation than that contained in the sum of the classes within it. One can expect additional variation between the classes, since the classes contain also structural variation (Leydesdorff, 1995).

The structural dimensions of the system may initially (and partially) be latent for the agents involved, but as the networks develop by further aggregating, the architecture of a social system can become more apparent (Lazarsfeld & Henry, 1968). A perceptive analyst is able to develop hypotheses about the latent dimensions of a virtually operating structure. The inference by the micro-constructivist that one would be unable to specify 'organizational principles or systems principles' accounting for the activity of individuals is no longer valid from this perspective.

The organizational principles can be explained historically in terms of how they have been constructed at the network level. They are not given naturally, but constructed historically. However, once constructed the constructions may begin to feed back on agency in a mode very similar to Giddens's 'duality of structure.' As against Giddens, however, the focus in Luhmann's constructivism is not on the construction process, but on what is constructed, that is, the social system. Note that our knowledge about this system has the status of a hypothesis. The social system should not be reified; it remains operational and under (re)construction.

Analytical theorizing about this operation can be informed by historical observations, but the systems theoretician takes a reflexive turn. In my opinion, Luhmann's sociology should primarily be read as a theory that informs sociological hypotheses by structuring them into a coherent framework. Any knowledge claim, however, remains itself an operational part of the social system, that is, as another account (Latour, 1988). From this perspective, sociological theorizing can be considered as contributing to empirical research by providing knowledge claims or hypotheses to be validated. As Luhmann (2002a, at p. 75) formulated it himself:

> The soundness of this reflection, however, arises—and this can still be ascertained by this reflection—from a form of social differentiation that no longer allows for any binding, authoritative representation of the world in the world or of society within society.

## Differentiation and integration

Accounts by participants to the membership contain an address different from accounts of sociological observers who wish to contribute to the development of their discipline. The social system differentiates in terms of roles. Neither 'the system' nor 'the situation' (nor 'everyday language' or 'action') necessarily integrates the different (sub)systems. 'Integration' is a special case that requires explanation. In a pluriform society, one expects frictions among discourses (based on observations and reports from different perspectives). The expected frictions can be observed, for example, in the case of competing paradigms.

The structural consequences of previous actions and interactions build up over time. People are historically constrained and enabled by structures that have been constructed at the supra-individual level. These structures are reproduced (or not) because of their institutionalized social functions. These insights about structure and function, of course, stem directly from Parsons. However, when Parsons's original 'unit of action' is replaced with 'interaction,' the systems under study are no longer only integrated by the operation. The interacting systems can both be integrated locally by action and at the same time differentiated in the reproduction. Interaction operates in cycles.

The cycles may begin to resonate. Different levels of nested interactions can be distinguished analytically. The levels can be considered condensations of these recursive operations, that is, communications about communications. For example, Luhmann (1975) distinguished between 'interaction,' 'organization,' and 'society.' The interacting agents can be expected to remain different, although they are able to exchange using an interface. When the networks reproduce distributions that are based on differences, the structural characteristics of these differences can be called 'differentiations.' The interfaces then also tend to become institutionalized, for example, as organizations.

Unlike symbolic interactionism with its pronounced focus on micro-level 'interactions,' Luhmann proposed to consider 'communication' as the basic operation of social systems. From this perspective, 'interaction' can be considered as a specific form of organizing communications, notably face-to-face communication in the present. As in Giddens's structuration theory, communications can also be aggregated and structured into contingent organizations and at the macro-level of society.

The starting point of this social systems perspective is that every action can also be considered as a communication among human beings (Luhmann, 1984, at p. 149). What cannot be communicated, cannot be considered as part of a social system. It should be observed that this definition includes non-verbal communication. Interaction is then the specific form of communication in which the participants are reflexively aware of the contingencies in the communication because of each other's presence. This 'double contingency' of the interaction structures action on both sides as a factor other than the individual lines of action. A structuration of the interactions can also be institutionalized, for example, in marriage. The reflexive awareness of the double contingencies and asymmetries in the mutual relations then induces cognition about the situation for each of the interacting actors.

From this reflexive perspective, action can be redefined—as in symbolic interactionism—as attributed by a network of social relations. However, the network perspective stands orthogonally to the actor perspective: agency is no longer considered as the cause and communication as the attribute, but vice versa, and the system is grounded in communication. An actor may take action (or not) given one's position in the network. The communication first provides the events happening with meaning. Meaning can be perceived by an actor (or not); reception is more crucial for interaction than taking action (Luhmann, 1990).

Providing 'meaning' to the events is crucial to all human and inter-human systems: social systems operate in terms of generating and reproducing meaning. This conclusion can be considered as a common heritage shared among Weber's sociology, Husserl's phenomenology, and the American pragmatists. Human beings interact reflexively, that is, in relation to one another; they evaluate whatever they observe, and although they are able to distinguish between the dimensions of 'facts' and 'values', the social science enterprise only takes off when the analysts also question what things mean to people.

The generation of 'meaning' at the social level can be considered as a consequence of human interaction. Individuals are able to entertain 'meaning' also discretionarily, but 'meaning' is reproduced by communication. Using a scheme from cybernetics, Luhmann then inverted the argument about the dynamics of meaning from the perspective of systems theory: human interaction can be reorganized by the social system of communications *because* social meaning is generated by interacting individuals. As meaning is repeatedly constructed bottom-up, the constructed (next-order) system tends to take over control when specific configurations can increasingly be stabilized.

Social systems and individuals can be expected to process meaning differently (Luhmann, 1986). For example, individuals can further develop as identities that may manage to map meaning one-to-one to their subjectivities. The axis for the representation at the social level, however, stands orthogonally to the axis of internal processing by an individual. Whereas the individual processes thoughts and consciousness, the social system enables us to develop, among other things, *discursive* knowledge.

### The generation of a knowledge base

When human beings interact, they generate uncertainty at the network level. One is able to handle this uncertainty because one has learned to cope with it by providing meaning to some actions and not to others. In the sociological literature, this has been discussed under the heading of the double contingency that provides meaning in social interactions (Parsons and Shills, 1951, at pp. 3-39; Parsons, 1968; Luhman, 1984, at pp. 148 ff.).

Both uncertainty and meaning can be expected to circulate among human beings. Languages enable us to provide a communication with meaning *and* to distinguish the expected information content of the message at the same time. This dual processing can be considered as the evolutionary achievement that has enabled the social system to develop the complex dynamics of a cultural evolution. The system develops in substantive and reflexive layers at the same time, but potentially in an uncoordinated way. The social system then emerges as a dynamic and flexible coordination mechanism among different levels of expectations.

The message provides meaning to the information contained in the message. For example, a word only has meaning in a sentence. Upon reception, the information can be rewritten as a signal of meaningful information and noise. This selective operation is recursive, that is, it can reflexively be applied upon itself. If the operation leaves traces over time, meaning can be invested in specific selections. The system can then provisionally be stabilized. In principle, stabilized systems can be further selected for globalization, that is, the historically achieved meaning can be compared with a horizon of possible meanings. Knowledge can then be developed as a next-order reflection allowing us to distinguish between meanings that make a difference and those which can provisionally be discarded as too uncertain. Thus, socially organized knowledge production further codifies the meaning-processing systems at a next-order layer.[2]

The stabilization of discursive knowledge in the social subsystem of scientific communication can be considered as a cultural achievement of the Scientific

---

[2] Does this mean that syntax would drive semantics? In a complex dynamics, the subdynamics (of syntax and semantics) do not drive each other, but co-produce the resulting phenomena by disturbing and constraining each other. The relative contributions of the subdynamics to the manifestations can vary situationally and over time.

Revolution of the 17th century. Individual knowledge production is then made interactive and in need of validation by communication. Modern sciences can no longer be understood in terms of the knowledge of single individuals. The study of the development of the sciences in terms of scientific revolutions (Kuhn, 1962) has made us aware of the nature of social systems of communication as different from individual consciousness systems and their sum totals (Leydesdorff, 2001a). The social system contains surplus value based on the interactions among human beings and their aggregates into groups.

The social system of interhuman expectations is initially nothing other than a plastic medium in which individuals process meaning and uncertainty, for example, by exchanging in these dimensions. When repeated over time the process can become increasingly structured. The media of social communication can become differentiated. A modern society, for example, is highly structured in terms of carefully constructed balances between different types of communication.

The operating structure of the social system is reproduced at the level of the social system by using our individual contributions as a variety of inputs (e.g., knowledge claims). But the processing is highly structured. Thus, human beings are not external to the system, but 'structurally coupled' to it in terms of the distribution of their inputs (Maturana & Varela, 1980). The network of communications can be expected to drift into provisional solutions of the puzzle of how to communicate all these inputs in an efficient (albeit perhaps suboptimal) way. The individual contributions provide the variation, while communication structures select by reinforcing some variations and not others.

The development of cognition as a next-order layer on each side—that is, at both the social level and within individuals—provides meaning-processing systems with another selective device that can feedback on lower-level selections and underlying variations. However, this mechanism is structured in social systems—Luhmann uses here the word 'dividuum'—differently from in individuals. While individuals process cognition internally, the social system manages to construct—under the historical conditions of emerging modernity—discursive knowledge as a control system for (scientific) communications.

**Functional differentiation of the communication**

What does the social system add when the inputs are selected for organizing the communications? At this end, the sociologist can build on metaphors available since the founding fathers of the discipline (e.g., Comte), notably, that the social system can be expected to develop evolutionarily in stages. First, there was the primitive organization of society based on kinship relations. This can be considered as a segmented system. Next, civilizations were formed based on a hierarchical and stratified structuration of the processing of meaning. In this stage, the levels of organization provided the main differentiation. However, the one-to-one correspondence between levels and control functions can be dissolved under historical conditions.

When the organization of society could no longer be contained within a single hierarchy (at the end of the Middle Ages), another format was gradually invented in the social system, notably, that of functional differentiation. This new form was shaped in the 15th century, for example, when the House of Burgundy ruled over the Low Countries. The Dukes of Burgundy were neither Emperors (of Germany) nor Kings (of France) and, therefore, they suffered from a lack of religious legitimation for claiming autonomy. Given the social and power relations of the time, monetary unification was invented as a means to bind their 'Empire of the Middle' (between France and Germany) together.

Philip the Good unified the monetary systems of Flanders, Brabant, Holland, and Hainault in 1433. In 1489, the silver 'stuiver' (or 'sous') was legally standardized as one twentieth of a golden guilder (florin) in all the Burgundian Netherlands (Groustra, 1995). This monetary union lasted until 1556. The coordination eroded because of the inflationary import of silver from the Spanish colonies during the 1540s and the protestant uprisings in the Netherlands in the 1550s. When the Dutch revolt gained momentum in the 1570s and 1580s, the northern provinces also decided that they no longer needed a King 'by the Grace of God,' but that they could organize the political system as a republic. The sciences and the arts, once set free from religious control, could then begin to flourish. The principle of functional differentiation entails that various symbolically mediated communication systems can operate to solve problems in society in a heterarchical mode, that is, alongside each other. Over time, these parallel systems can develop functionality for one another. Functionality, however, is further developed along orthogonal dimensions. Thus, one can expect that it will take time to develop from the stage of a breakdown of the horizontally stratified hierarchy into differentiation with functions along orthogonal axes as another mode of social organization.

The different function systems use various codes for providing meaning to the communication. Whereas the hierarchical (catholic) system had only a single center of control—that was based on a holy text—economic exchange relations, for example, could now be handled by making payments. The symbolically generalized medium of money makes it no longer necessary to communicate by negotiating prices verbally or imposing them by force. The specification of a price as an expected market value speeds up the economic transaction processes by organizing the communication in a specific (that is, functionally codified) format.

Functional differentiation first had to be invented and then also accepted as a solution to coordination problems at the level of the social system, for example, by recognizing privacy (e.g., in love relations) vis-à-vis public relations, market relations for exchange, and political state formation as different domains of communication. After the 'phase transition' from a hierarchical mode of communication to one in which functionality prevailed, the differentiation began to feed back on the institutional organization of society, for example, by questioning the functionality of the traditional organization. This was then reflected in an emerging discourse (during the 18th century and notably in France) about desirable forms of social organization.

Luhmann has emphasized in a series of studies entitled *Gesellschaftsstruktur und Semantik* ('The Structure of Society and Semantics') that although the semantic reflection is needed for stabilizing the functional differentiation, functional differentiation of communication should not be considered as a process within language, but one that precedes language structurally, that is, at the level of society. The communication becomes functionally differentiated as a *social* order; the semantic reflection and codification can be expected to lag behind. This social process of changing the mode of organizing communications among human beings can be expected to take centuries, and it cannot fully be completed because the complex system builds upon subdynamics that contain and reproduce forms of less complex organization as their building blocks.

For example, the hierarchical order of the communication in language with only Latin and then French as the *lingua franca*, was gradually replaced with a segmented system of 'natural languages' which could exist alongside each other as more or less equivalent. A system of nation states emerged in the 19$^{th}$ century as a sustainable form of shaping institutional structures that reintegrated the different functions in specific forms of organization. The prevailing tendency towards functional differentiation, however, is continuously upsetting the historical arrangements. Functional differentiation allows for handling more complexity at the global level since it is based on a next-order reconstruction. The reconstruction transforms all 'natural' (given) forms by infusing them with knowledge-based inventions. The global system, however, is constrained in terms of the development of retention mechanisms that enable its reproduction.

## The evolutionary mechanism

The American and French revolutions can perhaps be considered as the first deliberate attempts to reorganize a society institutionally so that it would be able to sustain the pluriform multiplicity of functions that characterize a modern society. The functional domains (e.g., markets, sciences) can be considered as global subsystems of communication, but at lower levels specific formats had to be generated in order to optimize the processing of information and meaning locally. While 'interaction' occurs also spontaneously between people, organizations have to be constructed.

Under the condition of functional differentiation, three levels can be distinguished at which one can expect that the function systems are recombined (cf. Luhmann, 1975): (i) in 'interaction' as face-to-face communication; (ii) organization in a social system provides criteria to distinguish those who are within from those external to a specific domain; and (iii) society can be considered as the coordination mechanism among functions at the global level. These three levels reconstruct segmentation, hierarchical stratification, and heterarchical differentiation of meaning processing, respectively. The organization of integration in institutions is thus analytically distinguished from the ongoing processes of functional differentiation among the globalizing subsystems (such as the economy and the sciences). The interfaces make possible translations among codes that provide different meanings to communications. But the interfaces have first to be invented and developed at specific places.

The functional subsystems operate by coding the communications specifically: for example, the market codes in terms of prices and payments, the sciences code communications in terms of whether they can be considered as functional for truth-finding and puzzle-solving, and political discourses code communication in terms of whether power and legitimation can be organized. Intimate relations code in terms of love and affection. The integrating levels, however, are not specific in terms of what is being coded. They solve the puzzle of how to interface the differences in codings locally. Agents at these different levels of aggregation can be expected to contribute to the differentiation by translating among differently coded meanings.

A range of global functions can be expected to resonate in inter-human communication. Functional differentiation means that some dimensions can be selected and others deselected in specific orders of communication. The integrating mechanisms can be considered as functional for organizing the differentiated communications at lower levels. They serve the retention of previously achieved levels of sophistication in the communication—or they may fail to do so. If they repeatedly fail to do so, an organization can be dissolved and replaced, yet without seriously affecting the dynamics of functional differentiation that can be expected to prevail at the global level.

This theorizing would remain completely speculative if it were not possible to develop empirical research questions on its basis. The historical example of how a monetary standard was developed at the end of the Middle Ages, provided us above with a first example of how one can use this theory as a heuristics for studying evolutionary developments in social processes. But can we also apply these methods more quantitatively and analytically? (Leydesdorff & Oomes, 1999)

What does a communication system do when it communicates? It selects a system's state for a communication. A social communication system can be expected to contain a very large number of system states, since the number of possible states increases with the number of the carrying agencies in the exponent. For example, if one throws two dice, one has $6^2$ (= 36) possible combinations. $N$ dice would provide us with $6^N$ possibilities, and similarly a group of ten people with six media for communication would allow for $6^{10}$, or more than 60 million possible combinations. A communication actualizes one or a few of these possibilities.

A large number of the actualizations may be volatile. One communication follows upon another without necessarily leaving traces. Selections then remain juxtaposed or, in other words, they are not correlated. However, selections may become correlated (if only by chance) in two respects, notably at the same moment in time and over the time axis. Along the time axis, 'variation' can be considered as change in relation to stability in the selections. At each moment in time, 'variation' can be considered as the sum of local disturbances, whereas structure selects for the function of this input. Structure, however, has to be built up historically before it can act as a systematic selector. In summary, a stabilized (and therefore observable) system contains two types of selections that operate concurrently: one by the network at each moment in time, and another over the time axis.

Note that the medium of communication thus provides us with a first constraint. When written communication is available to a social system, additional mechanisms of transmission become possible other than interactions and signaling in the present (Meyrowitz, 1994). Writing, however, has to be historically invented. As long as a communication system is mainly based on direct interaction, the span of communication is limited, and the selected states of the systems remain mainly juxtaposed. This can be recognized as a segmented order of social communications.

Writing is highly associated with the introduction of a new mode of control of communication, notably, the stabilization of a civilization (Innis, 1950). The mechanism of written communication enables cultures to span time periods at the supra-individual level and thus to stabilize systems of communication. Because the communication can also be saved for considerable periods of time, the new communications can be correlated to older ones and the selection of specific system states can be stabilized. Time breaks the symmetry in the mutual selections of a coevolution. Over time some previous selections can be selected for stabilization. In the phase space of possible selections the system then begins to develop along a trajectory. The shape of the historical trajectory is contingent upon the selections that the system manages to handle structurally. For example, a social system in which one is only able to write on clay tablets can be expected to develop differently from a social system in which papyrus or parchment have been invented. The relations between hierarchical interaction (command structures) and face-to-face interactions will vary among systems that are differently mediated.

Within civilizations based on hierarchies, the top of the hierarchy may be a king or an emperor with divine attributes. However, the prevalence of communication in the command structure can also become reflected. The invention of a holy text (e.g., the Bible) that integrates the system at a level more abstract than the physical presence of an emperor or king changes the cosmology. A civilization based on a more abstract set of principles can be considered as a high culture. But the reliance on communication—instead of physical force—as the basis for control is self-defeating in the long run because the constructed order needs to be enforced and the communicated order can then be recognized reflexively as historically specific.

The invention of new dimensions for the communication that can also be codified at the social level can be expected to turn the tables sooner or later (Arthur, 1988 and 1989). When the social system gains an additional degree of freedom, the new dimension allows the communicators to evade the dilemma of the two previously competing orders. At the edges of the spheres of influence between the Pope and the Emperor, for example, in city-states in Northern Italy and in the envisaged 'Empire of the Middle' in northern Europe one could develop trade, art, and sciences. The new communication structures would eventually challenge the catholic order spanning a single universe and its corresponding cosmology. The new order of communications can endure different dimensions of communication developing next to each other as different structures. Thus, the system recombines the advantages of segmented and stratified communication by inventing the mode of functional differentiation. Functional differentiation entails that communications can be distinguished with reference to the function of the communication. This provides new dimensions that were not available in a high culture.

For example, when the Netherlands were invaded by the French army in 1672, the Prince of Orange needed legitimation for the upcoming negotiations. He sent for Spinoza to join his cortège in order to impress the French generals. That Spinoza had been banned by the jewish and protestant churches for religious reasons was not in the interest of the Prince. In a functionally differentiated society, the representatives of functions can tolerate moderate conflict because the social system is no longer expected to process a single solution.

An order among the various function systems can be selected and reconstructed in a next round of reflections. If this additional degree of freedom can also be stabilized, this process globalizes the functional differentiation of the system. Some (provisional) stabilizations can be selected for globalization. *Globalization*, however, does not imply that a global system physically and/or meta-physically 'exists.' The functions refer only to a supersystem for which the subsystems can *analytically* be made functional. Initially, the existence of this supersystem remained a religious assumption; for example, Descartes' belief in the Goodness of God (*Veracitas Dei*) which would prevent Him from deceiving us all the time. Religious constructions like Leibniz' *harmonie préétablie* would guarantee a cosmological order in the universe.

Since the social system, however, continuously fails 'to exist' at the global level in a strong (physical or biological) sense—it remains a system of expectations—the organization of society can be expected to operate with the tensions between functional differentiation and locally organized integration of the communication. In the $18^{th}$ century, the Constitution was invented as an presumably 'universal' text that would bind all communicating agencies as members of a nation state. Soon, it became clear that each nation would have to develop its own constitution. The constitutions organized institutional systems of checks and balances that enabled the political economy to further develop on the basis of the level of functional differentiation that was achieved in the first half of the $19^{th}$ century. From this perspective, nations can be considered as institutional arrangements that include and exclude on the basis of nationality. In terms of evolution theory, they can be compared with niches. In ecology, niches are functional for the retention because they reorganize the complex environment by stabilizing boundaries.

When the system of nation states was completed (by approximately 1870), the national systems contained mechanisms for solving the major tensions between the state and civic society so that the function systems could be integrated locally, yet in a competing mode. From 1870 onwards, the social system has developed a new dimension to further improve these 'national' solutions. This new dimension can with hindsight be characterized as organized knowledge production and control (Whitley, 1984).

## Technological developments as inter-system dependencies

The sciences have developed continuously since the Scientific Revolution of the $17^{th}$ century (Price, 1961), but the fully developed political economies of the $19^{th}$ century provided the sciences with an institutional basis for further development. When the disciplines and the specialties then differentiated among themselves and in relation to their social contexts, the idea of a single and universal science had gradually to be abandoned. Interfaces with private appropriation by entrepreneurship and public control in science and technology policies were increasingly developed. Within science, the proliferation of disciplines and specialties made it possible to dissolve the idea of a single 'truth' to be discovered by science. One could proceed to a mode of 'truth-finding' and empirical 'puzzle-solving' (Simon, 1969). Thus, the code of scientific communication became internally differentiated (Gibbons *et al.*, 1994; Leydesdorff, 2001b).

Can the function systems also differentiate and complexify in terms of their interactions? In his 1990 study entitled *Die Wissenschaft der Gesellschaft* (The Science of Society), Luhmann formulated on p. 340:

> The differentiation of society changes also the social system in which it occurs, and this can again be made the subject of scientific theorizing. However, this is only possible if an accordingly complex systems theoretical arrangement can be specified (translation, LL).

Is the post-modern order thus eroding the system of functional differentiation (Sevänen, 2001)? When studying the so-called 'techno-sciences' as interface organizations with their own dynamics, one leaves the model of functional differentiation behind (Callon, 1998). Algorithmic models are needed which allow for next-order effects that are neither intended nor expected. Technological trajectories and regimes (Dosi, 1982), for example, can then be considered as endogenous consequences of non-linear interactions at the interfaces between the sciences ('supply') and markets ('demand').

Luhmann (2000, at p. 396) has discussed the organization of interfaces as structural interruptions of the communicative order at the global level. As he formulated it:

> Society has to develop beyond functional differentiation and use another principle of systems formation in order to gain the ultrastability and therefore sufficient local capacity to absorb irritations by providing organization.

What might this ultra-stabilization of an interaction between functionally differentiated sub-systems mean? Stability requires a form of integration by organization. Indeed, an important condition for the development of modern high-tech sciences seems to be the increasing integration of political, economic, and scientific orientations in research practices (Gibbons *et al.* 1994).

Professional practices can be considered as organized interaction systems that allow for specific recombinations of integration and differentiation in new roles. Integration in the sense of de-differentiation, however, would be evolutionarily unlikely, since the social system might then lose its capacity to handle complexity. Thus, these constructed interaction systems are heavily organized, but from the perspective of interactions.

Alternatively, the constructions can be shaped at the level of interactions among organizations. For example, technological developments can be considered as the result of inter-systemic resonances which have been stabilized as new functions in the social system during the last century. The stabilization of interfaces and the *discursive construction of integration* can then be considered as instances of an emerging next-order of global communications.

This higher-order communication can be expected to contain a new *epistèmè* (Foucault, 1972 [1969], at p. 191): in addition to the communication of substantive novelty and methodologically warranted codification ('truth'), high-tech sciences, for example, are translating representations of subsystems of society into scientific knowledge by modeling them, and *vice versa*, by legitimating research results in 'trans-epistemic' cycles of communication (Knorr-Cetina, 1982 and 1999). In other words, one is institutionally warranted in changing the code of the communication, for example, because of a flexible division of labour within the research community.

The emerging patterns of the high-tech sciences are not expected to replace the older models, but to encompass them and to guide their future development. The next-order regime entrains the trajectories on which it builds (Kampmann *et al*., 1994). In other words, 'high tech' and 'big science' can be considered as results of an 'epistemic drift' of translations between economic innovations and research questions; and *vice versa*, of the possibility to merge fundamental and applied research questions in terms of selections of relevant representations (Elzinga, 1985 and 1992). These newly emerging communication systems contain more than a single codification, and additionally they are able to translate between these codifications internally by using a spiral model of communication. Using computer simulations, for example, developments can be analyzed in terms of processes of representation and communication within relevant scientific-political-economic communities (Ahrweiler, 1995): high-tech sciences develop by communicating in terms of recursive selections on interactively constructed representations.

In my opinion, the emergence of 'big science' and patterns of international collaborations in science during the second half of the 20th century can be considered as the institutional acculturation of the new *epistèmè.* The reflexive reorganization of these institutional patterns by using new forms of S&T policies was apparently delayed until the second oil crisis of 1979, when the post-war system entered into a serious crisis at the level of the global economy. The gradual development of stable patterns of scientific reproduction in fields like 'artificial intelligence', 'biotechnology', and 'advanced materials' in the 1980s and 1990s indicates the viability of a new mode of scientific communication.

### The globalization of the knowledge base of expectations

The local networks of institutions like universities, industries, and governments can be considered as carriers of a next order of potentially global communications. These systems can then be expected to go through a phase transition in terms of their need for new communicative competencies. The translations no longer occur between 'natural' languages, but between functional codes of communications that are themselves entrained in a flux. This next-oder system emerges *within* the system as its globalization. The existence of a global system, however, remains a hypothesis. Since this hypothesis is entertained and communicated, the global level potentially restructures the expectation structures in the globalizing systems. By being transformed on this basis, all 'naturally given' or 'historically constructed' bases of underlying systems tend to become increasingly 'knowledge-based'.

Remember that some selections were selected for stabilization along the time axis. By globalizing the system entertains the time axis no longer as a historical symmetry-breaking mechanism, but as another dimension. The local realization can then be evaluated from a global, that is, knowledge-based perspective. The global perspective operates on the present state of the system by enabling us to entertain the idea that what has historically been constructed 'ain't necessarily so.' In other words, it can always be reconstructed on the basis of new insights.

A knowledge-based system operates on the basis of the current state of the system as one of its possible representations. Each historical representation can be compared to others. The global perspective *adds* an expectation to the local perspectives. Since the various perspectives compete for the explanation of what can be expected to occur, neither the global nor the local perspective can claim priority. The perspectives remain analytically juxtaposed (as hypotheses!), but they interact. As noted, this means at the level of 'organization' that the single organization is increasingly networked and that the inter-institutional arrangements become more important for the functionality of organization than the single perspective.

What does this globalization of the knowledge-base mean for 'interaction'? It seems to me that this can already be observed, for example, in the form of the role of e-mail communication as an addition to previously existing forms of interaction. We have increasingly become aware that interaction is mediated and that one can entertain various forms of interaction with different objectives. Furthermore, one is increasingly able to anticipate interactively the unintended consequences of previous communications. Interactions can thus be expected to become increasingly recognizable as translations among differently coded communications.

The programmatic view of symbolic interactions that interactions can only be concatenated bottom-up in order to inform us about social structure can then no longer be maintained without running into serious problems. The methodological restrictions of micro-constructivism have practical implications. Interactions are situated, and thus next-order levels of nested interactions and communications can be expected to resonate within the observables. The situation is overdetermined by expectations based on hypothetical structures. The systems theoretical program in sociology adds and informs the hypotheses about the feedback loops within the interactions that it studies.